\documentclass[preprint]{emulateapj}
\usepackage{hyperref}
\usepackage[usenames,dvipsnames]{color}
\hypersetup{colorlinks,citecolor=Blue,linkcolor=Red,urlcolor=Blue}
\usepackage[titletoc]{appendix}
\usepackage{amsmath}
\usepackage{float}
\newcommand{\be}{\begin{equation}}
\newcommand{\ee}{\end{equation}}
\def\lta{\,\raise 0.3 ex\hbox{$ < $}\kern -0.75 em
 \lower 0.7 ex\hbox{$\sim$}\,}
\def\gta{\,\raise 0.3 ex\hbox{$ > $}\kern -0.75 em
 \lower 0.7 ex\hbox{$\sim$}\,} 

\begin{document} 

\title{A secular resonant origin for the loneliness of hot Jupiters} 

\author{Christopher Spalding$^1$ and Konstantin Batygin$^1$} 
\affil{$^1$Division of Geological and Planetary Sciences\\
California Institute of Technology, Pasadena, CA 91125} 
\affil{$\,$}

\begin{abstract}
Despite decades of inquiry, the origin of giant planets residing within a few tenths of an astronomical unit from their host stars remains unclear. Traditionally, these objects are thought to have formed further out before subsequently migrating inwards. However, the necessity of migration has been recently called into question with the emergence of in-situ formation models of close-in giant planets. Observational characterization of the transiting sub-sample of close-in giants has revealed that ``warm" Jupiters, possessing orbital periods longer than roughly 10\,days more often possess close-in, co-transiting planetary companions than shorter period ``hot" Jupiters, that are usually lonely. This finding has previously been interpreted as evidence that smooth, early migration or in situ formation gave rise to warm Jupiter-hosting systems, whereas more violent, post-disk migration pathways sculpted hot Jupiter-hosting systems. In this work, we demonstrate that both classes of planet may arise via early migration or in-situ conglomeration, but that the enhanced loneliness of hot Jupiters arises due to a secular resonant interaction with the stellar quadrupole moment. Such an interaction tilts the orbits of exterior, lower mass planets, removing them from transit surveys where the hot Jupiter is detected. Warm Jupiter-hosting systems, in contrast, retain their coplanarity due to the weaker influence of the host star's quadrupolar potential relative to planet-disk interactions. In this way, hot Jupiters and warm Jupiters are placed within a unified theoretical framework that may be readily validated or falsified using data from upcoming missions such as \textit{TESS}.
\end{abstract}

\section{Introduction}

\begin{figure*}[ht]
\centering
\includegraphics[trim=0cm 0cm 0cm 0cm, clip=true,width=1\textwidth]{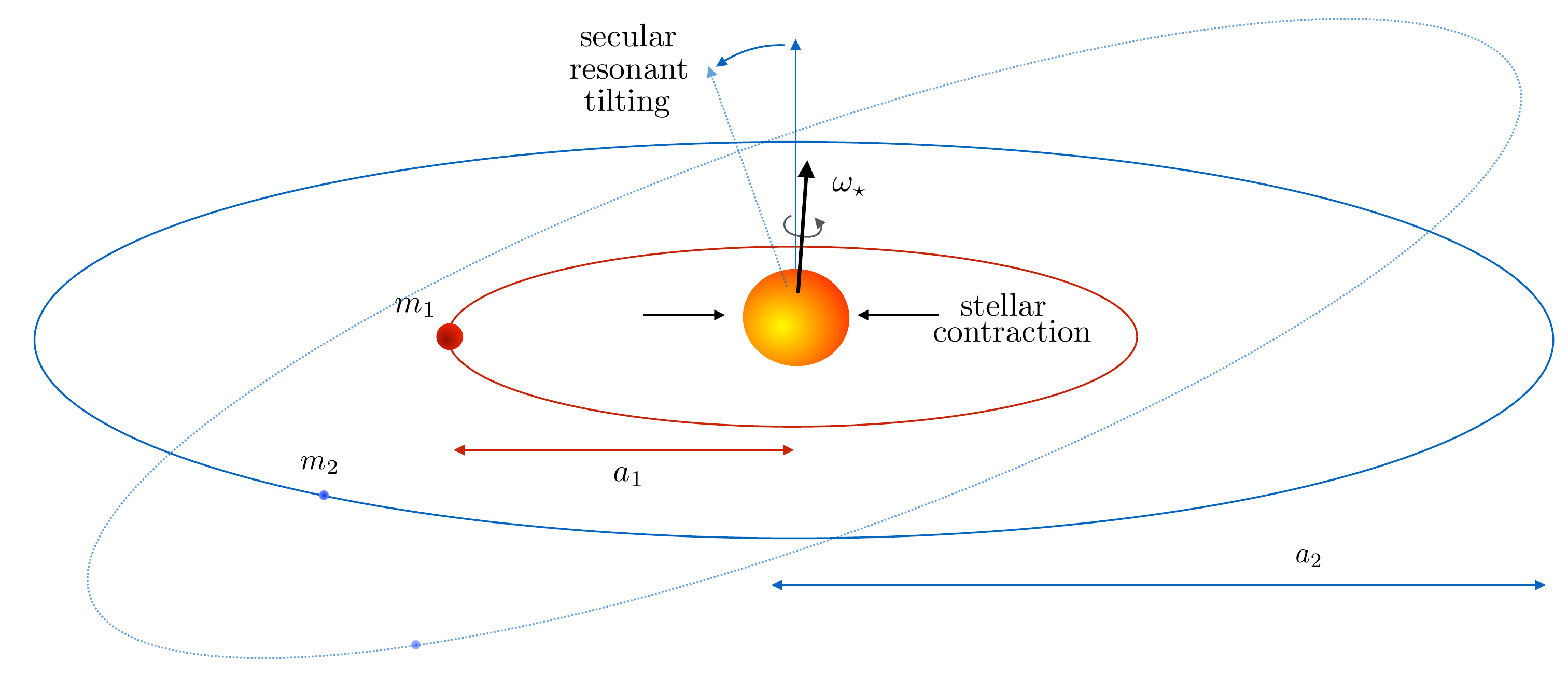}
\caption{A schematic of the set up considered in the text. A giant planet with mass $m_2$ follows a circular orbit with semi-major axis $a_1$. Exterior, lies a lower-mass planet $m_2$ on circular orbit with semi-major axis $a_2$. The exterior orbit is forced to undergo nodal regression due to a combination of stellar quadrupolar potential and secular perturbations from the inner giant planet's orbit. The giant's orbit, in turn, is regressing mostly owing to secular perturbations form the stellar oblateness. Initially, the inner planet regresses faster but as the stellar quadrupole decays (as a result of physical contraction), a commensurability is encountered in the two frequencies, leading to the secular resonant excitation of mutual inclinations between the planets.}
\label{Schematic}
\end{figure*}

Arguably the longest-standing problem in exoplanetary science concerns the origin and evolution of so-called ``hot Jupiters" \citep{Mayor1995}. Planets in this category are loosely defined as possessing masses comparable to Jupiter, but residing on orbits with periods shorter than about 10 days. Similarly to the giant planets in our solar system, these objects are thought to have formed through the ``core-accretion" pathway \citep{Stevenson1982,Pollack1996}. Within this framework, a $\sim10-15$ Earth mass solid core conglomerates whilst still embedded within the natal protoplanetary disk, accretes a comparable-mass envelope and subsequently initiates a period of runaway gas accretion, yielding a Jupiter-mass planet. 

Traditionally, the in situ formation of hot Jupiters has been considered impossible, owing to the difficulty in constructing sufficiently large cores within the hot inner regions of the protoplanetary nebula, where solid grains are relatively scarce \citep{Lin1996,Rafikov2006}. Consequently, the prevailing notion is that hot Jupiters formed further out, beyond the snow line, before subsequently migrating inwards. This migration can occur either during the disk-hosting stage (through so-called type II migration; \citealt{Lin1996,Kley2012}) or later, via the excitation of large eccentricity followed by tidal circularization \citep{Wu2003,Beauge2012}. For brevity, we will group these migration mechanisms into two categories: ``early" for when the hot Jupiters reach their close-in orbits before the disk disperses, and ``late," referring to migration proceeding subsequent to disk-dispersal.

A separate sub-population of giant planets that is progressively becoming better characterized is the ``warm Jupiter" class of close-in bodies \citep{Steffen2012,Huang2016}, which are defined, again loosely, as residing on orbits of period $\sim10-30$\,days. Like the hot Jupiters, these objects lie interior to the ice line, and therefore suffer from many of the same arguments against in situ formation as their hotter counterparts. However, a crucial difference is that warm Jupiters exhibit tidal circularization timescales that are typically too long to have migrated via a late pathway, particularly when no exterior giant companion is detected \citep{Dong2013,Petrovich2016}. Accordingly these giants appear to have attained their close-in orbits prior to disk dispersal.

Within published literature, the distinction between ``warm" and ``hot" has been somewhat arbitrary. However, the recent study of \citet{Huang2016}, along with some earlier investigations \citep{Steffen2012}, have pointed out an empirical distinction between the transiting subsample of the two populations. Specifically, the fraction of warm Jupiters possessing close-in (i.e., periods $<50$\,days), co-transiting companions is roughly 50\%, in contrast to the hot Jupiters, where the analogous fraction is close to 0\%, with only one counter-example, WASP-47b currently known \citep{Becker2015,Weiss2017}. This pattern has been interpreted as evidence in support of a high-eccentricity origin for hot Jupiters, with close-in companions being cast out during the migration. In contrast, warm Jupiters were interpreted to arise from smooth migration within a natal disk, where WASP-47b constitutes the innermost tail of this population.

Recent work has begun to question the necessity of migration for the formation of close-in giant planets \citep{Boley2016,Batygin2016}. In particular, \citet{Batygin2016} considered the long-term dynamical evolution of a close-in giant planet forming in situ, with a super Earth residing on an exterior orbit. As the host star contracts, and as the giant planet grows, the outer planet transitions from a regime where its nodal regression rate is smaller than that of the giant's, to a regime where the two frequencies are approximately commensurate. This results in a convergent encounter with a secular resonance that tilts the orbit of the outer companion, potentially all the way to 90$^\circ$. Moreover, if the giant planet's orbit is sufficiently eccentric ($e\gtrsim0.05$) the outer orbit may instead have its eccentricity raised, by a similar secular resonance in terms of the precession of perihelia, leading to dynamical instability within the system. From the point of view of transit surveys, both outcomes will lead to a lonely close-in giant, except for fortunate viewing geometries.

A key limitation of the above picture, mentioned in \citet{Batygin2016}, is that tidal interactions with the disk gas itself may induce nodal regression upon the outer planet's orbit that quenches the resonant tilting that would otherwise occur in the absence of a disk \citep{Hahn2003}. In other words, if the physical mechanism responsible for the onset of the secular resonance occurs whilst the disk is still around, the system may retain coplanarity, and the giant planet will co-transit with its close-in companions. For the purposes of this work, ``in situ" formation is dynamically equivalent to ``early" migration, because both processes lead to systems that are already close-in at the time of disk dissipation, and so we make no statements regarding which of the two scenarios is more likely.

The key finding of this paper is that the resonance is encountered later for closer-in systems. That is, hot Jupiter-hosting systems encounter the resonance later than warm Jupiter systems. Given fiducial disk lifetimes and stellar rotation periods \citep{Haisch2001,Bouvier2013}, we expect giant planets to become lonely when orbiting interior to $\sim0.1\,$AU (or, $\approx11.6$\, days for a solar mass star, i.e., suggestively close to the warm Jupiter - hot Jupiter divide). At larger orbital radii, systems may still encounter the resonance, but will do so whilst embedded in the disk gas, which can therefore prevent inclination excitation. 





 In brief, we show that hot Jupiters will become lonely in transit surveys, not because they formed differently from warm Jupiters, but because they encountered the aforementioned resonance after their disk dissipated, when nothing prevented their outer companions from being driven to high inclinations.

\section{Analytical Theory}

To set up the problem, suppose a giant planet, of mass $m_1$ orbits at a semi-major axis $a_1$ interior to a less massive planet, say, a super Earth of mass $m_2$ at semi-major axis $a_2$ (Figure~\ref{Schematic}). Young stars possess large radii $R_\star$ and rotate rapidly \citep{Shu1987,Bouvier2013}, leading to significant oblateness. This oblateness, parameterized via the second gravitational moment $J_2$, leads to precession of the argument of perihelion $\varpi$ and regression of the longitudes of ascending node $\Omega$ of the planetary orbits. 

The protoplanetary disk is expected to damp inclinations and eccentricities to small values \citep{Kley2012}, allowing the gravitational disturbing potential influencing the planets to be expanded to second order in eccentricities and inclinations (or 4th order near secular resonances \citep{Batygin2016}). At such an order, the inclination and eccentricity degrees of freedom are decoupled and therefore may be treated in isolation. In this work, we assume the orbits to be circular, such that $\Omega$ and inclination $i$ together constitute the only degree of freedom. Alternatively one may assume the orbits are coplanar, but instead consider the eccentricity degree of freedom in isolation. The logic and numerical coefficients are very similar in both cases, so we do not work through both, but include a brief discussion of eccentricity dynamics in Section~3.1. 

\subsection{Stellar Evolution}

For a circular orbit with semi-major axis $a_{\textrm{p}}$ and mean motion $n_{\textrm{p}}$, the stellar-induced nodal regression rate is given by \citep{Murray1999}
\begin{align}\label{nu}
\dot{\Omega}_{\textrm{p}}\equiv\nu_{\star,\textrm{p}}=\frac{3}{2}n_{\textrm{p}}J_2\bigg(\frac{R_\star}{a_{\textrm{p}}}\bigg)^2,
\end{align}
where $p=\{1,\,2\}$; see Figure~\ref{Schematic}. The aforementioned second gravitational moment, $J_2$ is related to the stellar spin angular velocity $\omega_\star$ and Love number (twice the apsidal motion constant) $k_2$ through the relation \citep{Sterne1939}:
\begin{align}\label{Jaytwo}
J_2=\frac{1}{3}\bigg(\frac{\omega_\star}{\omega_b}\bigg)^2\,k_2,
\end{align}
where $\omega_b$ is the break-up angular velocity of the star. The Love number can be extracted from polytropic stellar models with index $\chi=3/2$ (i.e., fully convective; \citealt{Chandrasekar1939}), leading to $k_2\approx0.28$ \citep{Batygin2013}, which is the numerical value we adopt throughout. The spin periods of T Tauri stars may be constrained through observation \citep{Bouvier2013} and in general lie within the range of $\sim1-10$\,days. Substituting expression~(\ref{Jaytwo}) into Equation~(\ref{nu}) yields,
\begin{align}\label{simplified}
\nu_{\star,\textrm{p}}=n_{\textrm{p}}\frac{k_2}{2} \bigg(\frac{\omega_\star}{n_{\textrm{p}}}\bigg)^2\bigg(\frac{R_\star}{a_{\textrm{p}}}\bigg)^5,
\end{align}
thereby casting the nodal regression rate in terms of quantities that are either directly observable or may be inferred from simple models.

With time, the central star will contract ($\dot{R}_\star<0$) and so $\nu_{\star,\textrm{p}}$ will decrease. Additional time-dependence may arise owing to stellar spin-down, however, owing to the high order of $R_\star$ in equation~{\ref{simplified}} we ignore any changes in $\omega_\star$ (as discussed below, the dynamics do not depend sensitively on the details, only that $\nu_{\star,\textrm{p}}$ decreases). 

To a good approximation, the contraction of a protostar may be modelled as Kelvin-Helmholtz contraction of a polytropic body \citep{Batygin2013}: 
\begin{align}\label{Rstar}
R_\star(t)=R_{\star,0}\bigg(1+\frac{t}{\tau_\textrm{c}}\bigg)^{-\frac{1}{3}},
\end{align}
where we define the contraction timescale (1/3 of the Kelvin-Helmholtz timescale);
\begin{align}
\tau_{c}\equiv\frac{G M_\star^2}{28 \pi \sigma T_{\textrm{eff}}^4 R_{\star,0}^3}.
\end{align}
In this expression, $G$ is the gravitational constant, $\sigma$ is the Steffan-Boltzmann constant and the stellar mass $M_\star$ is taken equal to 1 solar mass ($M_\odot$) throughout. The above analytic form agrees well with the numerical pre-main sequence evolution models of \citet{Siess2000} provided a value of $R_{\star,0}\gtrsim6R_\odot$ is chosen\footnote{The initial radius simply has to be large, rather than an exact value because the star essentially loses information about initial conditions within the disk's lifetime.} for a solar mass star with surface temperature $T_{\textrm{eff}}=4270$\,K. 


\subsection{Capture into secular resonance}

Planet-planet interactions will induce modal regression in addition to that arising from the stellar quadrupole. It can be shown through linear secular perturbation theory \citep{Murray1999,Morby2002} that the time-averaged regression rate of the inner planet $<\nu_1>$ takes the form
\begin{align}\label{nu1}
<\nu_1(t)>\approx \nu_{\star,1}(t)+\frac{1}{4}\frac{m_2}{M_\star}\bigg(\frac{a_1}{a_2}\bigg)^2 b_{3/2}^{(1)}\bigg(\frac{a_1}{a_2}\bigg)n_1,
\end{align}
where the first term results from the oblateness of the host star, which decays with time, and the second term arises owing to planet-planet interactions. Furthermore, we have introduced the function $b_{3/2}^{(1)}(a_1/a_2)$, known as a Laplace coefficient \citep{Murray1999}, defined as:
\begin{align}
b_{3/2}^{(1)}(\alpha)\equiv\frac{1}{\pi}\int^{2\pi}_0\bigg[\frac{\cos(\psi)}{(1+\alpha^2-2\alpha\cos \psi)^{3/2}}\bigg]d\psi.
\end{align}
Analogously, the time-averaged nodal regression rate of the outer planet is given by
\begin{align}\label{nu2}
<\nu_2(t)>\approx \nu_{\star,2}(t)+\frac{1}{4}\frac{m_1}{M_\star}\bigg(\frac{a_1}{a_2}\bigg) b_{3/2}^{(1)}\bigg(\frac{a_1}{a_2}\bigg)n_2.
\end{align}

We will assume that the semi-major axes and masses are fixed, reflecting the fact that most planet-building and migration occurs before the disk dissipates. Then, the ratio between stellar and planetary orbital angular momenta is given by
\begin{align}
j\equiv\frac{I_\star M_\star R_\star^2 \omega_\star}{m_p\sqrt{G M_\star a_p}}
\end{align}
where $I_\star\approx0.21$ is the dimensionless moment of inertia, as calculated for a fully convective, polytropic star \citep{Chandrasekar1939}. For nominal hot Jupiter parameters; a Jupiter-mass planet at $0.05$\,AU, orbiting a star with $R_\star\sim2 R_\odot$, spinning with a period of 3\,days, we find that $j\gg1$. As a result, we will consider the stellar spin axis to be fixed, a statement that the planetary orbits possess significantly less angular momentum than the star \citep{Batygin2016}. We note that this approximation begins to break down as the star contracts, the stellar rotation period is long, or if the planet is situated further out, i.e., in the warm Jupiter regime. Whereas a full account of the star spin dynamics is not expected to affect our arguments with respect to planet-planet inclinations, numerous recent observational investigations \citep{Li2016,Dai2017} are beginning to detect a trend whereby more distant transiting planets exhibit larger spin-orbit misalignments. While suggestive, we shall not explore this aspect of the problem further here. 

\begin{figure}[h!]
\centering
\includegraphics[trim=0cm 0cm 0cm 0cm, clip=true,width=1\columnwidth]{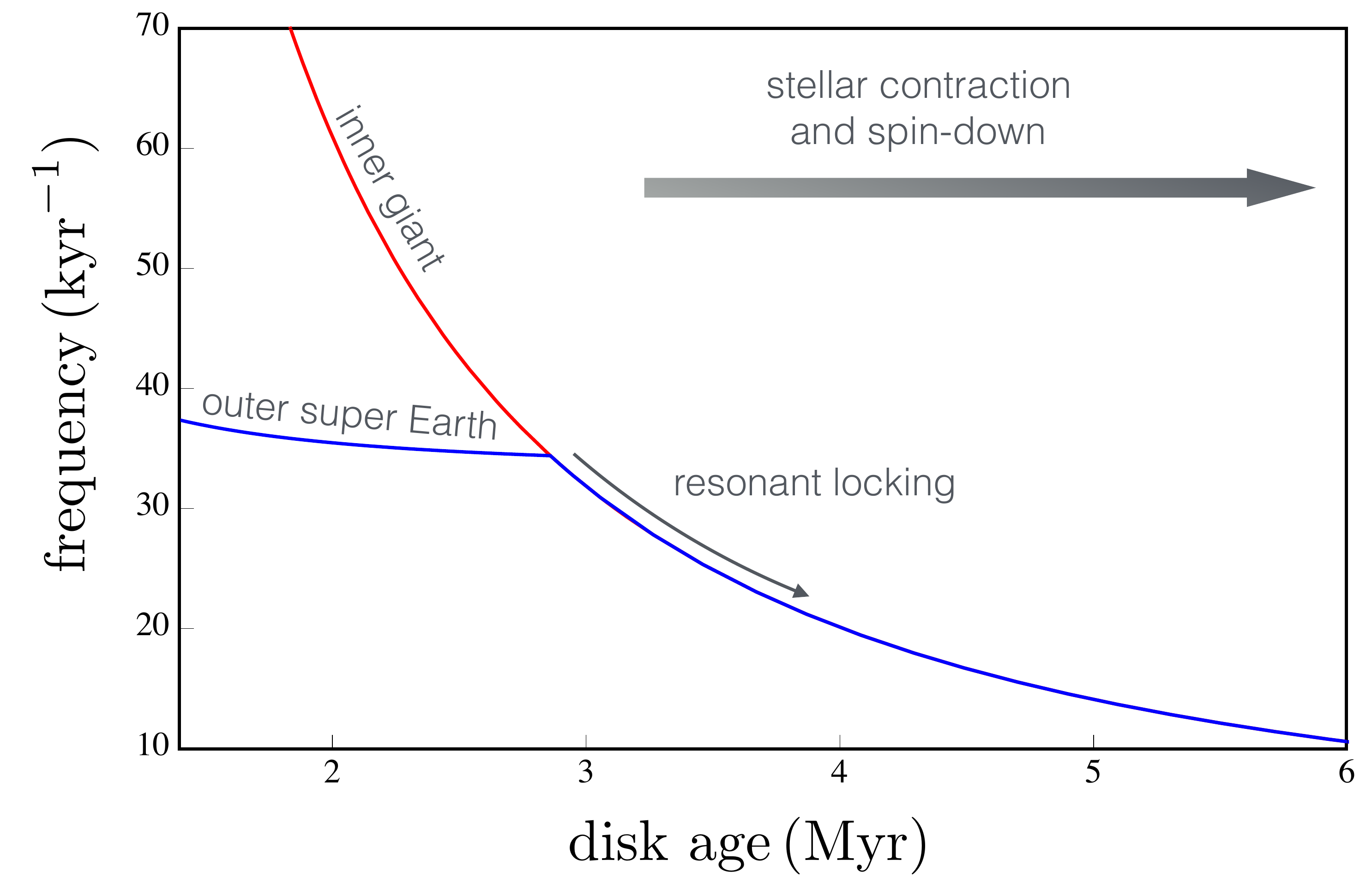}
\caption{The time evolution of the nodal regression frequencies for both planets as the host star contracts. The requirement to turn a giant planet-super Earth system into an apparently lonely giant is that $\nu_2<\nu_1$ (i.e., the red line is above the blue line) at the point when the disk dissipates, such that a point is crossed where the two frequencies are roughly commensurate. As argued in the text, this will always happen as the giant grows, but can be bypassed due to planet-disk interactions. If this is the picture dominating the hot Jupiter-warm Jupiter distribution, we would expect to see more hot Jupiters with companions around faster-rotating, massive stars and a gradual drop in companion fraction toward smaller semi-major axes. Parameters used in this illustrative figure are $m_1/M_\star=10^{-3},\,m_2/M_\star=10^{-5},\,a_1=0.04\,\textrm{AU},\,a_2=0.1\,\textrm{AU},\,P_\star=3\,\textrm{days}$ around a solar mass star. Resonance is encountered at $t=2.86\,\textrm{Myr}$.} 
\label{nus}
\end{figure}

 The stellar quadrupole's influence upon the inner planet will be significantly larger than its effect upon the outer planet ($(\nu_{\star,1}/\nu_{\star,2})=(a_2/a_1)^{7/2}\gg1$). Consequently, as the star contracts, the inner planet's nodal regression frequency will decrease more rapidly than that of the outer planet's. For nominal parameters, the initial state has $\nu_1>\nu_2$, but at a later stage the star has contracted to the point where the two frequencies become similar (Figure~\ref{nus}). As discussed in more detail in \citet{Batygin2016}, the result of such a commensurability in frequencies is capture into a secular resonance and excitation of inclination within the outer companion's orbit. Accordingly, if the giant was detected through transit it would appear ``lonely" except under a fortuitous viewing geometry.  
\newpage
 \subsection{Influence of the disk}

 The above discussion showed how close-in giants may become lonely, but made little distinction between ``hot" and ``warm" Jupiter-hosting systems. In this section, we show that considering gravitational interactions between the planets and their natal disk naturally leads to closer-in systems experiencing resonant excitation of inclinations more often than more distant systems, thereby accounting for the increased loneliness of hot vs warm Jupiters.
 
  Determining the detailed dynamics of planets embedded within gaseous disks constitutes an active field of research on its own \citep{Kley2012}. Here we will concern ourselves with a simplified yet instructive description of the dynamics, involving two qualitatively separate processes. Specifically, planet-disk interactions tend to damp the eccentricities and inclinations of embedded planets \citep{Tanaka2004}, together with inducing precession in the planetary arguments of perihelion and regression of the longitudes of ascending node \citep{Hahn2003}. 
 
 We will show below that the most important effect for our purposes is the disk-induced regression of the node. Tidal disk-planet interactions are expected to damp the outer planet's orbital inclination over a timescale given approximately by \citep{Tanaka2004}
\begin{align}
\tau_{\textrm{inc}}\equiv\bigg|\frac{di_2}{dt}\frac{1}{i_2}\bigg|^{-1}\approx \zeta\frac{P_2}{2\pi} \bigg(\frac{M_\star}{m_2}\bigg)\bigg(\frac{M_\star}{\Sigma_2 a_2^2}\bigg)\beta^4,
\end{align}
where $\beta\equiv h/a_2\approx0.05$ is the disk's aspect ratio, $\Sigma_2$ is the disk's surface density at $a_2$, the numerical constant $\zeta\approx 2$ and $P_2$ is the outer planet's orbital period.

In order to determine whether the disk's inclination damping will inhibit the adiabatic growth of inclination described above, we introduce the libration timescale for the resonant argument $\Omega_1-\Omega_2$ within resonance \citep{Morby2002,Batygin2016}
\begin{align}
P_{\textrm{lib}}\approx\frac{P_2}{2\pi}\frac{2}{3s_1}\bigg(\frac{M_\star}{m_1}\bigg)\bigg(\frac{a_2}{a_1}\bigg)^2,
\end{align}
where $s_1\equiv\sin(i_1/2)$, with $i_1$ being the inclination of the inner planet's orbit with respect to the stellar spin axis. The secular resonant inclination excitation will be prevented if the adiabatic limit is broken \citep{Henrard1993}, i.e., 
\begin{align}\label{crit}
\tau_{\textrm{inc}}\lesssim P_{\textrm{lib}}.
\end{align}
If we suppose the disk to follow a minimum mass solar nebula profile \citep{Hayashi1981} of $\Sigma(a)\approx 2000(a/1\textrm{au})^{-3/2}$g\,cm$^{-2}$, $a_2\approx 0.2$\,AU, $a_1=0.1$\,AU, $m_2=10^{-5}\,M_\odot$ and $m_1=10^{-3}\,M_\odot$ the disk will damp inclination growth provided $s_1\lesssim 0.2$, or $i_1\lesssim20^\circ$. Accordingly, disk-induced inclination damping may prevent the secular resonance from driving inclination growth for systems aligned with their natal disk. However, owing to the uncertainties in both the calculation of damping timescales and the order-of-magnitude nature of inequality~\ref{crit}, it is difficult to determine whether resonant growth is prevented in all cases.  

Given the uncertainties in computing the importance of direct disk-driven inclination-damping, let us now consider an additional mechanism by which the disk may prevent resonant growth of inclination; the regression of nodes induced by the disk's gravitational potential. It may be shown that the disk's quadrupole induces a regression rate of \citep{Hahn2003}
\begin{align}
\nu_{pd}&\approx n_p\frac{\pi \Sigma_p a_p^2}{M_\star \beta},
\end{align}
and the modal regression occurs in the same direction as that induced by the stellar quadrupole and planet-planet interactions. If one assumes a surface density profile of the MMSN, $\Sigma_p\propto a_p^{-3/2}$ \citep{Hayashi1981} the disk-induced modal regression frequency $\nu_{pd}\propto a_p^{-1}$, such that more distant planets experience a slower rate arising from the disk's quadrupole. 

In the interest of simplicity, and for illustrative purposes, we only include the disk's effect upon the outer, smaller planet. Ultimately our arguments are not particularly sensitive to this decision. However, giant planets are expected to open gaps in the disk \citep{Crida2006}, and the MRI-active inner region of the nebula is expected to possess a lower gas density \citep{Armitage2011}. Both of these effects are likely to significantly diminish the disk's secular influence upon the inner giant's orbit. Including the disk-induced nodal regression to $<\nu_2(t)>$ given above yields
\begin{align}
\big<\big(\nu_2(t)/n_2\big>\big)&\approx\frac{3}{2}J_2\bigg(\frac{R_\star}{a_2}\bigg)^2\nonumber\\
&+\frac{1}{4}\frac{m_1}{M_\star}\bigg(\frac{a_1}{a_2}\bigg) b_{3/2}^{(1)}\bigg(\frac{a_1}{a_2}\bigg)+\frac{\pi\Sigma_2 a_2^2}{\beta M_\star}.
\end{align}

%

\begin{figure*}[th]
\centering
\includegraphics[trim=0cm 0cm 0cm 0cm, clip=true,width=1\textwidth]{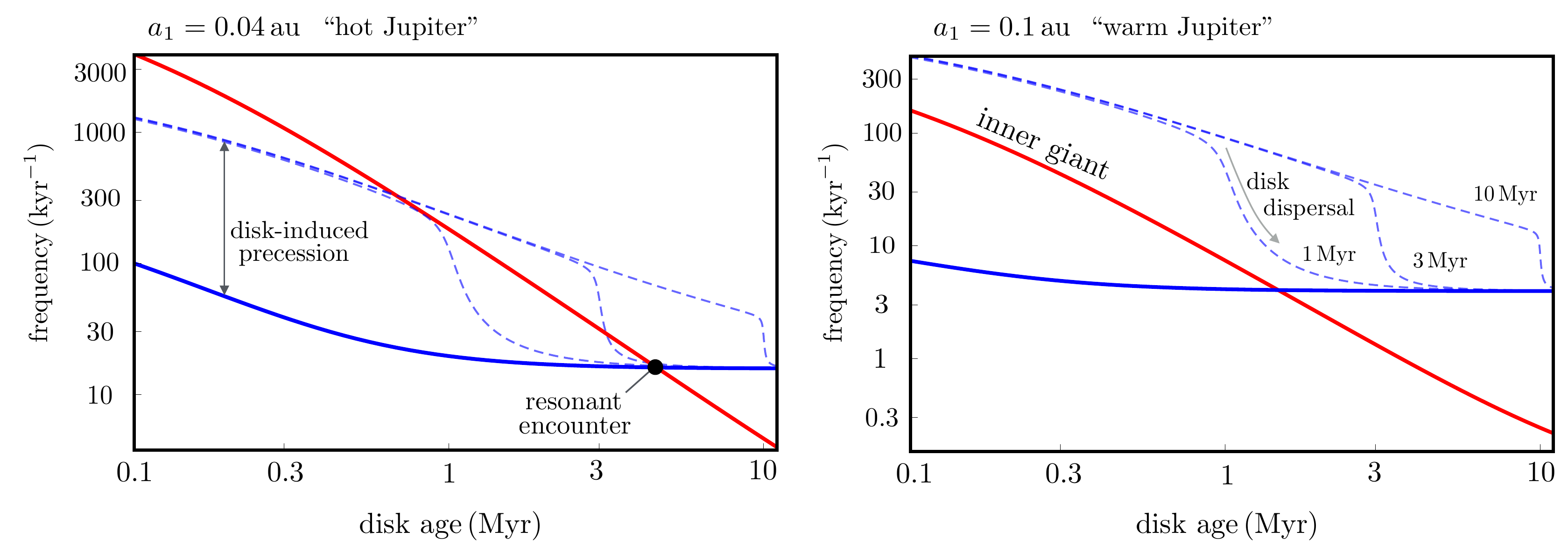}
\caption{An illustration of the influence of the disk's quadrupole. The solid blue line represents the outer planet's nodal regression in the case with no disk, whereas the three dashed lines represent the case where the disk's quadrupole moment is included. When the outer planet is forced to regress faster than the giant (red line) throughout the entire disk lifetime, the secular resonant encounter described in the text is prevented. The left panel considers a hot Jupiter, at 0.04\,AU interior to a test particle at 0.12\,au. The right panel depicts the case for a Warm Jupiter at 0.1\,AU with an exterior test particle at 0.3\,au. The closer, hot Jupiter system encounters the secular resonance later (black circle) and so the disk is more likely to have dispersed, whereas the warm Jupiter system entirely bypasses the secular resonance even for very short disk lifetimes (e.g., 1\,Myr).}
\label{DiskDamp}
\end{figure*}

In order to determine the importance of disk-induced nodal regression, we must formulate the evolution of the disk's mass and surface density. Observational constraints upon disk masses remain challenging, and intrinsic variation between the evolution of individual disks makes a single, generalized parameterisation impossible. For our purposes, it suffices to use an average disk mass evolution. The observed decay of disk accretion rates with time \citep{Calvet2005} is well approximated by the following parameterization
\begin{align}\label{diskMass}
M_{\textrm{disk}}(t)=\frac{M_{\textrm{disk,0}}}{1+t/\tau_v},
\end{align}
where the viscous decay time $\tau_v=0.5$\,Myr and the initial disk mass $M_{\textrm{disk,0}}=0.05$M$_\odot$. We may sanity-check this result by writing the surface density profile as 
\begin{align}
\Sigma(a,t)=\Sigma_0(t)\bigg(\frac{a}{a_0}\bigg)^{-3/2}
\end{align}
yielding the disk mass 
\begin{align}
M_{\textrm{disk}}(t)=4\pi\Sigma_0(t=0)a_{\textrm{out}}^{1/2}a_0^{3/2}.
\end{align}
In order to achieve $M_{\textrm{disk,0}}=0.05$\,$M_\odot$ the disk must begin with a surface density at $1$\,AU of $\Sigma_0(t=0)\approx6500$g\,cm$^{-2}$, which is similar to the ``minimum-mass extrasolar nebula" inferred by requiring the observed populations of close-in super Earths formed close to their current locations \citep{Chiang2013}. Accordingly, prescription~(\ref{diskMass}) constitutes a reasonable approximation to the disk's global evolution.

Viscous evolution alone is notoriously unable to match observational deductions regarding disk dispersal timescales. Rather, the current consensus is that, after $\sim1-10$\,Myr of viscous evolution, the disk disperses over a short ($\Delta t\lesssim10^5$\,years) timescale \citep{Alexander2014}, beginning with the inner few au. To account for the two-timescale nature of disk evolution we parameterize the surface density as
\begin{align}
\Sigma_0(t)=\frac{\Sigma_0|_{t=0}}{1+t/\tau_v}\bigg[\frac{1}{2}-\frac{1}{\pi}\arctan\bigg(\frac{t-\tau_d}{\Delta t}\bigg)\bigg].
\end{align}
Where $\tau_d$ is the time of disk dispersal, ranging from $\sim1-10$\,Myr.

In the disk-free case (Figure~\ref{nus}) $<\nu_2>$ is always greater than $<\nu_1>$ initially. However, the influence of the disk upon the outer planet increases its regression frequency above that of the inner giant when the system is relatively far from the central star. To illustrate this effect, we present the evolution of $<\nu_1>$ and $<\nu_2>$ for two systems (Figure~\ref{DiskDamp}). The first is a hot Jupiter, orbiting at $a_1=0.04$\,AU interior to a companion at $a_2=0.12$\,au. In this scenario, the disk quadrupole is not sufficient to take the outer planet's regression rate above the giant's and so the resonant criterion is met at a disk age under 1\,Myr. 

If the the giant planet has already formed at this early stage, secular resonant excitation will proceed. However, if the giant is not yet formed, the system has a second chance to encounter resonance, after the disk disperses. When the disk dispersal time is $\tau_d=1$\,Myr or $\tau_d=3$\,Myr, the system encounters secular resonance as it would have in the absence of a disk (Figure~\ref{nus}). Owing to the strong influence of the central star's quadrupole, only very long-lived disks will disperse after the system can subsequently encounter secular resonance (e.g., 10\,Myr in this case) and so in most cases, this hot Jupiter will end up appearing lonely in transit surveys.
 
   The second case presented would generally be described as possessing a ``warm Jupiter" at 0.1\,AU and an exterior companion at 0.3\,au. In this case, the influence of the disk overcomes the nodal regression induced by the more distant central star's oblateness. Accordingly, the resonance is quenched and inclination excitation does not occur. If, for example, the disk had dissipated already by $\sim1$\,Myr, this system would undergo resonant capture and become a lonely warm Jupiter system. More typical disk lifetimes \citep{Haisch2001} tend to prevent secular resonant tilting when the giant planet lies beyond $\sim0.1$\,AU - which is close to the oft-quoted dividing line between what is considered a ``warm" or a ``hot" Jupiter.

\subsection{Criterion for Loneliness}

The above discussion shows that when the disk is still present, it is likely to quench the secular resonant dynamics, especially in more distant systems. Consequently, the criterion for whether a giant planet will resonantly excite the inclination of its outer companion is simply that \textit{the inner planet is regressing faster than the outer planet at the time of disk dissipation}:
\begin{align}
\boxed{\nu_1\big|_{t=\tau_{\textrm{d}}}>\nu_2\big|_{t=\tau_{\textrm{d}}}}.
\end{align}
By substituting expressions~(\ref{simplified}, \ref{Rstar}, \ref{nu1} and \ref{nu2}) into the above inequality, we may reformulate the criterion in terms of a critical stellar spin period at the time of disk dissipation $P_\star\big|_{t=\tau_{\textrm{d}}}$below which inclination is excited. After some algebra, we arrive at the criterion for tilting to occur and the generation of a lonely giant planet:
\begin{align}\label{criterion}
P_\star\big|_{t=\tau_{\textrm{d}}}&\lesssim P_1\sqrt{2k_2}\bigg(\frac{M_\star}{m_1}\bigg)^{1/2}\bigg(\frac{R_\star(\tau_{\textrm{d}})}{a_1}\bigg)^{5/2}\nonumber\\
&\times\Bigg[\frac{1-\alpha^{7/2}}{\alpha^{5/2}b_{3/2}^{(1)}(\alpha)}\Bigg]^{1/2}\bigg(1-\frac{\Lambda_2}{\Lambda_1}\bigg)^{-1/2},
\end{align}
where $P_1$ is the inner planet's orbital period, $\alpha\equiv a_1/a_2$ and $\Lambda_{\textrm{p}}\equiv m_{\textrm{p}}\sqrt{GM_\star a_{\textrm{p}}}$. 

It is worth noting that the above criterion was derived within the secular regime, and so assumes that the two planets are not caught in first or second order mean-motion resonances. Terms associated with these resonances enter the disturbing potential at first and second order respectively in eccentricity, potentially swamping the second order secular dynamics considered here \citep{Murray1999}. Accordingly, if our criterion is to be applied to close-in giant planet systems with observed exterior companions, care must be taken to check whether the resonant arguments appropriate to first or second order mean motion resonances are librating or circulating. In general, for nearly circular orbits, this is equivalent to requiring that the period ratios lay more than a few per cent from exact commensurability, though precise libration widths depend upon the resonance considered and the eccentricity.

\section{Results \& Discussion}

\begin{figure}
\centering
\includegraphics[trim=0cm 0cm 0cm 0cm, clip=true,width=1\columnwidth]{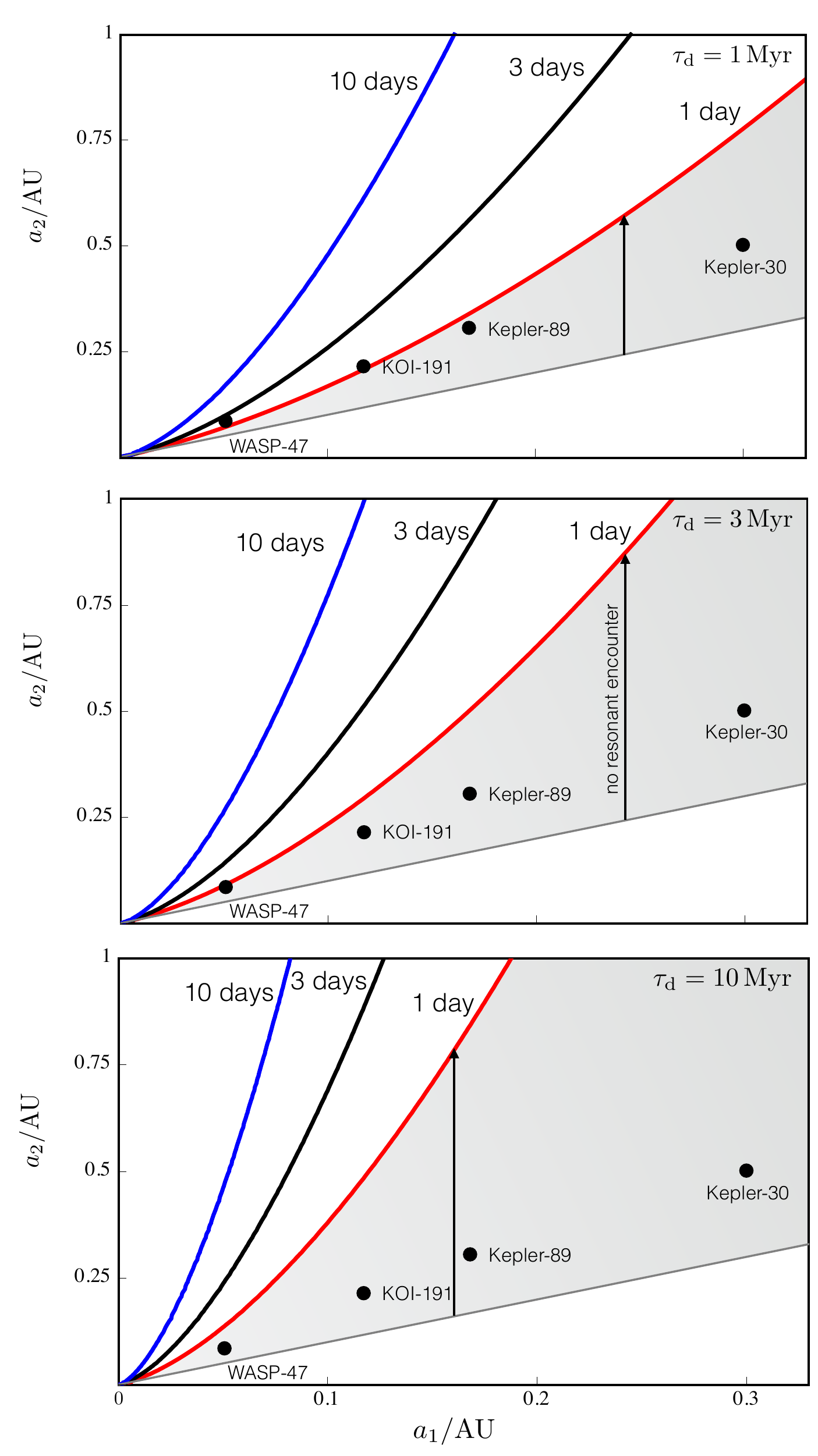}
\caption{Outer companions within the shaded region will not encounter the resonance and will remain coplanar with the inner giant. We have plotted the configuration of the four systems known where a giant lies interior to a lower-mass planet. All four lie within the region where coplanarity is expected to persist around a well-aligned star. Interestingly, the innermost example, WASP-47, lies almost exactly on the boundary, consistent with it being the closest-in known example and only hot Jupiter with a close outer companion.}
\label{FriendZone}
\end{figure}

The criterion~(\ref{criterion}) is represented in Figure~\ref{FriendZone} by way of 3 plots, corresponding to 3 nominal times of disk dissipation, $\tau_{\textrm{d}}=\{1\,\textrm{Myr},\,3\,\textrm{Myr},\,10\,\textrm{Myr}\}$. Using the above prescription for stellar contraction (equation~\ref{Rstar}), these times correspond to stellar radii $R_{\star}(\tau_{\textrm{d}})=\{2.4\,R_{\odot},\,1.7\,R_{\odot},\,1.1\,R_{\odot}\}$. Within each plot, we display lines representing 3 stellar spin periods $P\big|_{t=\tau_{\textrm{d}}}=\{10\,\textrm{days},\,3\,\textrm{days},\,1\,\textrm{day}\}$, spanning the observed range of T Tauri spin periods \citep{Bouvier2013}, with $m_1=10^{-3}M_{\odot}$, $m_2=10^{-5}M_{\odot}$ and $M_\star=M_{\odot}$. For the relevant spin period and $\tau_{\textrm{d}}$, an outer planet with $a_2$ above the line is expected to encounter resonance and become misaligned. The region where coplanarity is expected to be maintained even for a relatively fast 1\,day stellar rotation period is shaded.

The key message of Figure~\ref{FriendZone} is that there is significantly more semi-major axis space available for close, outer companions to Jupiters residing beyond $\sim0.1$\,AU than hotter Jupiters. Therefore, the loneliness of closer-in giants may naturally arise owing to the greater tendency for their outer companions to become resonantly inclined and taken out of transit surveys. 

Superimposed on Figure~\ref{FriendZone}, we have placed points representing the four cases in \citet{Huang2016} where a giant planet lies interior to a close companion. All four lie within the shaded region, consistent with our hypothesis. The hot Jupiter system WASP-47 might have misaligned WASP-47d if its disk dissipated early ($\sim 1\,$Myr) and the star was particularly rapidly-rotating ($\sim1$\,day; see top panel). As more examples are detected, we expect the shaded region to become filled in to a significantly greater extent than the regions between the 1 and 10\,day lines. 

\subsection{The Hot Jupiter - Warm Jupiter Distinction}

Despite the somewhat arbitrary distinction between a ``hot" versus a ``warm" Jupiter, the dearth of close-in companions to hot Jupiters has been taken as evidence that there is a physical distinction between the formation pathway of each planetary regime \citep{Steffen2012,Huang2016}. In this paper, we have shown that separate formation pathways are not required to explain the loneliness of hot Jupiters. Simply by forming closer to their host stars, these systems were more susceptible to perturbations from the host star's quadrupole moment, facilitating a secular resonance that tilted their outer companions. 

Our proposed mechanism is directly falsifiable in that we would not expect to find transiting outer companions above the upper-most line in Figure~(\ref{FriendZone}), though too large of an $a_2$ will eventually break our assumption that the star possesses most of the system's angular momentum, altering the criteria derived above somewhat. Additionally, owing to uncertainties in disk lifetime and T Tauri spin rates, finding a transiting outer companion between the 3\,day and 10\,day lines is not necessarily a falsification -- one can simply suppose the young star was rotating rapidly, or the disk was late to dissipate. Consequently, predictions may be falsified most readily in a statistical sense. 

In the framework of our model, fewer outer companions should be found to transit outside the shaded region in Figure~(\ref{FriendZone}) than in the shaded region, even after correcting for observational biases. The semi-major axis distribution within the shaded region is expected to resemble those of lower-mass \textit{Kepler} systems \citep{Tremaine2012,Morton2016}. Having only four examples makes it difficult to rigorously evaluate these hypothesis currently. 
 
We should note that it is possible to view mutually inclined orbits via transit if their lines of nodes are fortunately commensurate with the line of sight. Consequently, if all hot Jupiters possess inclined companions, we would expect to see a small fraction of those companions in transit surveys \citep{Steffen2012}. The lack of such detections suggests that any close companions to hot Jupiters are not simply misaligned, but lost from the system entirely. 

We propose two potential mechanisms whereby the dynamics described here will not simply misalign, but remove close companions to hot Jupiters. First, lower-mass \textit{Kepler} systems often exhibit high multiplicity \citep{Fabrycky2014}, as opposed to the simple 2-planet system described here. It has been previously demonstrated \citep{Spalding2016} that misaligning one or more components of such closely-packed systems has the potential to destabilize the entire architecture. Furthermore, any additional planets within the system introduce extra secular modes and potential resonances that the system sequentially encounters as the central star contracts. 

A second mechanism for complete loss of companions arises if the giant planet's orbit possesses an eccentricity higher than $e_1\sim0.05$. Here, an additional secular resonance becomes applicable whereby the outer orbit has its eccentricity, as opposed to its inclination, raised. This process will eject the outer planet from the system by way of a lowering of its pericenter until the orbits cross \citep{Batygin2016}. As alluded to in Section~2, the quantitative criteria for encountering this eccentricity resonance are similar to those of the inclination resonance \citep{Murray1999}, but disks are generally expected to damp planetary eccentricities to a point where the inclination resonance should dominate \citep{Kley2012,Batygin2016}. However, many warm Jupiters are known to be eccentric \citep{Dong2013}, potentially forced by exterior giant companions \citep{Bryan2016}, hinting that super Earths may occasionally be lost through the eccentricity resonance. Cumulatively, more work is needed to elucidate how often companions are expected to be tilted versus lost entirely.
 

\subsection{Inner Companions}

This work has focused on the tilting of outer companions to giant planets, and has talked little of inner companions. By inspection of equation~(\ref{criterion}), one can see that the secular resonance cannot be encountered if the inner planet has less angular momentum then the outer planet ($\Lambda_1<\Lambda_2$). Accordingly, the resonant misalignment mechanism proposed here is not capable of explaining the absence of inner companions to hot Jupiters versus warm Jupiters \citep{Huang2016}. However, the only known inner companion to a hot Jupiter, WASP-47e, resides at the particularly close-in distance of 0.017\,au, not much larger than the radius of the young star itself ($0.017\,\textrm{au}\approx3.6 R_\odot$). Whereas we have not identified a specific mechanism that removes inner companions, their rarity may simply arising owing to limited physical space within a hot Jupiter's orbit. We expect that more interior companions will be found by the \textit{TESS} mission \citep{Ricker2015}, however, they are likely to be rare.
 
  \section{Summary}


 The existence of giant planets inside the snowline has traditionally favoured an explanation whereby the planet itself, or at least its multiple Earth mass core, forms at large radii before subsequently migrating inwards. This migration can occur early, during the disk-hosting stage \citep{Kley2012}, or long after, through a high-eccentricity pathway \citep{Wu2003,Beauge2012}. Between the two formation channels, the high-eccentricity pathway encounters more theoretical challenges in forming warm Jupiters \citep{Dong2013}, and empirical challenges in forming hot Jupiters \citep{Dawson2014}. 
 
 With a single exception, hot Jupiters are found to possess no close-in transiting companions, in contrast to their slightly cooler counterparts, the warm Jupiters, roughly half of which are found with co-transiting close companions \citep{Steffen2012,Becker2015,Huang2016}. This evidence appears consistent with a high-eccentricity origin for hot Jupiters and an early, possibly in situ, origin for warm Jupiters. Interestingly, around 50\% of Super Earth systems are thought to be significantly inclined \citep{Johansen2012}, possibly also owing to mutual inclinations induced by the host star \citep{Spalding2016}. 
 
 Systems of super Earths are generally thought to form through early migration or in situ \citep{Lee2016}, much like warm Jupiter systems. Thus, in the picture of high-eccentricity migration, hot Jupiters are exceptional - the lone class of planets who swung in from the outer regions of the planetary system long after the dispersal of the protoplanetary disk. In this Paper, we have demonstrated that by taking account of the stellar quadrupole moment, hot Jupiters and warm Jupiters may have formed through identical pathways. However, the hot Jupiters, precisely because of their close-in configuration, encounter a secular resonance after the dispersal of their natal disk, and tilt their outer companions' orbits beyond the reach of transit surveys. 
 
 The relative unimportance of the Sun to the Solar System planets' orbits has caused many to ignore stellar non-sphericity in exoplanetary systems. In contrast, the results of this paper, and of related works (\citealt{Batygin2016,Spalding2016}), illustrate the key role of the stellar oblateness to the long-term dynamical evolution, and eventual architecture, of compact exoplanetary systems. 

\begin{acknowledgments}
 This research is based in part upon work supported by NSF grant AST 1517936 and the NESSF Graduate Fellowship in Earth and Planetary Sciences (C.S). We would like to thank the anonymous referee for comments that improved the quality of the manuscript.
\end{acknowledgments}

\end{document}